\documentclass[11pt,a4paper]{article}
\usepackage{amsfonts}
\usepackage{latexsym}
\usepackage{amsmath,amssymb}
\usepackage{mathrsfs}
\usepackage{cite}

\def\bbc{{\Bbb C}}
\def\bbr{{\Bbb R}}
\def\bbz{{\Bbb Z}}

\newtheorem{rem}{Remark}[section]

\newtheorem{ex}{Example}

\def\tr{\mathrm{tr\,}}

\def\ad{\mathrm{ad\,}}

\def\diag{\mathrm{diag\,}}
\def\rmi{\mathrm{i}}

\def\rmid{\mathrm{id\,}}

\begin{document}
\title{Multicomponent Nonlinear Evolution Equations of the Heisenberg Ferromagnet Type. Local
versus Nonlocal Reductions}
\author{Tihomir Valchev \\
\small Institute of Mathematics and Informatics,\\
\small Bulgarian Academy of Sciences,\\
\small Acad. Georgi Bonchev Str., Block 8, 1113 Sofia, Bulgaria}
\date{}
\maketitle

\begin{abstract}
The paper is dedicated to a system of matrix nonlinear evolution equations related to a Hermitian
symmetric space of the type $\mathbf{A.III}$. The system under consideration extends the $1+1$
dimensional Heisenberg ferromagnet equation in the sense that its Lax pair has a form rather
similar to the pair of the original Heisenberg ferromagnet model. We shall present here certain local
and nonlocal reductions. A local integrable deformation and some of its reductions will be discussed too.
\\[0.2cm]
\end{abstract}

\section{Introduction}\label{sec:1}

Let us consider the following matrix system
\begin{equation}
\begin{split}
\rmi\mathbf{u}_t + \left(\mathbf{u}_x\mathbf{v}^T\mathbf{u} - \mathbf{u}\mathbf{v}^T\mathbf{u}_x\right)_x &= 0,\\
\rmi\mathbf{v}_t - \left(\mathbf{v}_x\mathbf{u}^T\mathbf{v} - \mathbf{v}\mathbf{u}^T\mathbf{v}_x\right)_x &= 0 
\end{split}
\label{ghf_matgen}\end{equation}
where the subscripts denote partial differentiation in independent variables, the superscript $T$ means matrix
transposition and "$\rmi$" is the imaginary unit. The rectangular complex $n\times m$-matrices $\mathbf{u}(x,t)$
and $\mathbf{v}(x,t)$ are not independent but are required to fulfil the constraints
\[\mathbf{u}\mathbf{v}^T\mathbf{u} = \mathbf{u},\qquad \mathbf{v}\mathbf{u}^T\mathbf{v} = \mathbf{v}.\]
Then the system (\ref{ghf_matgen}) has zero curvature representation with Lax operators of the form
\begin{eqnarray*}
L(\lambda) & = & \rmi\partial_x - \lambda S, \qquad S = \left(\begin{array}{cc}
0 & \mathbf{u}^T \\ \mathbf{v} & 0
\end{array}\right),\\
A(\lambda) & = & \rmi\partial_t + \lambda A_1 + \lambda^2 A_2,\qquad 
A_1 = - \rmi[S^2,S_{x}],\\
A_2 & = & \frac{2r}{m+n} I_{m+n} - S^2, \qquad r = \frac{1}{2}\tr S^2.
\end{eqnarray*}
Above $\lambda\in\bbc$ is spectral parameter, $[\,.\, ,\,.]$ is the usual commutator, $I_{m+n}$ is the unit matrix
of dimension $m+n$ and $r$ is some positive integer ($0<2r<m+n$). In this aspect (\ref{ghf_matgen}) represents a formal
integrable generalization of the Heisenberg ferromagnet equation 
\begin{equation}
\rmi S_t = \frac{1}{2} [S,S_{xx}], \qquad S = \left(\begin{array}{cc}
S_3 & S_1 -\rmi S_2 \\ S_1 + \rmi S_2 & - S_3
\end{array}\right),\qquad S^2 = I_2
\label{hf}\end{equation}
that seems to be a novel one. The equation (\ref{hf}) is integrable through inverse scattering
transform \cite{blue-bible}. There are various integrable generalizations of the Heisenberg ferromagnet
equation known in literature, e.g. a generalization of (\ref{hf}) for higher dimensional unitary or special
linear algebras \cite{BoPo,forkul,yantih1,ntades17} or $2+1$-dimensional extensions, see \cite{ishim, mmnl}.

In this report we aim at presenting and discussing certain local and nonlocal reductions of the generic matrix
system (\ref{ghf_matgen}). Recently nonlocal reductions have become quite popular \cite{gerdji,ggi,gurpek} mostly 
due to the paper \cite{ablomus}. 

Another issue to be paid attention in the text refers to construction of a deformation of (\ref{ghf_matgen})
that preserves integrability. We introduce a local and a nonlocal reduction of the generic deformation. The
local reduction is a natural extension of that published in \cite{gmv} while the latter is written here for
the first time.

The report is organized as follows. Next section is dedicated to linear bundle Lax pairs in pole gauge
associated with Hermitian symmetric spaces of the type $\mathbf{A.III}$. We show that the zero curvature 
condition of those $L$-$A$ pairs produce integrable matrix equations of the Heisenberg ferromagnet
type. Section \ref{reds} contains certain reductions of those generic generalized Heisenberg
ferromagnet equations. In section \ref{deform} we introduce an integrable deformation of our generalized
Heisenberg ferromagnet equations. For that purpose we deform the Lax pair to another more complicated
rational bundle Lax pair related to the same type of symmetric spaces. Last section contains some concluding
remarks and comments.

\section{Linear Bundles Related to $\mathbf{A.III}$ Symmetric Spaces}

In this section we introduce a Lax pair associated with Hermitian symmetric spaces of the type
$\mathbf{A.III}$ and present some basic facts necessary for the considerations to follow. A more 
profound exposition on the general relation between linear bundle Lax pairs and symmetric spaces
can be found in \cite{forkul}. The zero curvature condition of these Lax operators lead to a generic
system of matrix nonlinear evolution equations (NEEs) of the Heisenberg ferromagnet type. 

Let us start with a few remarks on the notation to be used further in text. We are going to
write $M_{m,n}$ for the space of all $m\times n$ matrices with complex entries. Then
$\mathrm{SL}(n)\subset M_{n,n}$ will stand for the special linear group of degree
$n$ over complex numbers and $\mathrm{SU}(p,q)\subset\mathrm{SL}(n)$ will be the pseudounitary group ($p+q=n$).
The corresponding Lie algebras will be denoted by $\mathfrak{sl}(n)$ and $\mathfrak{su}(p,q)$
respectively. For any $X\in M_{m,n}$ we shall write $X^T$ and $X^{\dag}$ for the matrices that
are transposed to $X$ and Hermitian conjugated to $X$ respectively, i.e. we have
$\left(X^T\right)_{ij} = X_{ji}$, $\left(X^{\dag}\right)_{ij} = X^*_{ji}$ ($*$ stands for complex
conjugation). The unit matrix of dimension $m$ will be denoted by $I_m$ and the adjoint
operator of any square matrix $X$ will be defined through the usual commutator of matrices, i.e.
$\ad_X(\,.\,) = [X,\, .\, ]$ holds.
    
Let us introduce the Lax pair 
\begin{equation}
\begin{split}
L(\lambda) &= \rmi\partial_x - \lambda S,\qquad S(x,t)\in\mathfrak{sl}(m+n),\qquad\lambda\in\bbc, \\
A(\lambda) &= \rmi\partial_t + \lambda A_1 + \lambda^2 A_2,\qquad
A_{1,2}(x,t)\in \mathfrak{sl}(m+n),
\end{split}\label{lax}\end{equation}
where the potential function $S$ obeys the polynomial constraint
\begin{equation}
S^3 = S .
\label{s_cond}\end{equation}
We also require that the above $L$-$A$ pair is subject to 
\begin{equation}
HL(-\lambda)H = L(\lambda),\qquad HA(-\lambda)H = A(\lambda),\qquad
H = \diag(-I_m,I_n)
\label{h_red}\end{equation}
which means the Lax pair is related to the symmetric space $\mathbf{A.III}
\equiv \mathrm{SU}(m+n)\slash\mathrm{S}(\mathrm{U}(m)\times\mathrm{U}(n))$.
Equivalently one may say that the matrix coefficients of (\ref{lax}) satisfy the conditions
\begin{equation}
HS(x,t)H = - S(x,t) ,\quad HA_{\sigma}(x,t)H = (-1)^{\sigma}A_{\sigma}(x,t),\qquad \sigma = 1,2.	
\label{h_red_comp}\end{equation}  
The action of $H$ induces the $\bbz_2$-grading 
\begin{eqnarray*}
\mathfrak{sl}(m+n) &=& \mathfrak{sl}^0(m+n) + \mathfrak{sl}^1(m+n),\\
\mathfrak{sl}^{\varsigma}(m+n) &=& \{X\in \mathfrak{sl}(m+n)|\  HXH = (-1)^{\varsigma}X\},
\quad \varsigma  = 0,1
\end{eqnarray*}
in the Lie algebra. In view of the condition (\ref{h_red_comp}) we have that $S(x,t), A_1(x,t)\in\mathfrak{sl}^{1}(m+n)$
and $A_2(x,t)\in\mathfrak{sl}^{0}(m+n)$.

\begin{rem}
From the constraint (\ref{s_cond}) one may deduce that the spectrum of $S$ consists of $\pm 1$
and $0$. It is also straightforward to see that $S^2$ is a projector of rank that is an even integer to be
denoted by $2r$. The multiplicity of the eigenvalues $\pm 1$ and $0$ is $r$ and $m+n-2r$ respectively.
We shall require that $0<2r<m+n$ so that the trivial cases $S^2=0, I_{m+n}$ are excluded. Then from (\ref{s_cond})
it immediately follows that $S\neq 0$ which is a rather natural requirement.
\end{rem}

The compatibility condition $[L(\lambda), A(\lambda)] = 0$ of (\ref{lax}) leads to the equations
\begin{eqnarray}
&&[S,A_{2}] = 0, \label{l3}\\
&&\rmi\partial_x A_{2}  - [S, A_{1}] = 0, \label{l2}\\
&&\partial_x A_{1} + \partial_t S = 0. \label{l1}
\end{eqnarray}
The first recurrence relation above along with the conditions (\ref{s_cond}) and (\ref{h_red}) imply we can 
pick up $A_2$ in the form
\begin{equation}
A_2 = \frac{2r I_{m+n}}{m+n} - S^2,\qquad \tr S^2 = 2r.
\label{a2}\end{equation}
In order to find the matrix coefficient $A_1$ from (\ref{l2}), we split $A_1$ into a $S$-commuting part
$A_1^{\mathrm{s}}$ and some remainder $A_1^{\mathrm{r}}$. Obviously there is an infinite number
of such splittings differing from each other by some $S$-commuting term. This is why we shall specify the
splitting by requiring
\[\pi_{S}A_1^{\mathrm{s}} = 0, \qquad \pi_{S}A_1^{\mathrm{r}} = A_1^{\mathrm{r}},\]
where the projector $\pi_{S}=\ad^{-1}_S\ad_{S}$ is defined through the following right inverse
of the adjoint operator of $S$: 
\begin{equation}
\ad_{S}^{-1} = \frac{1}{4}\left(5\ad_{S} - \ad_{S}^3\right).
\label{adinv}\end{equation}
More details on this particular choice of $\ad_{S}^{-1}$ can be found in \cite{yantih1}. 

Taking into account (\ref{a2}) and  (\ref{adinv}), the relation (\ref{l2}) leads to  
\begin{equation}
A_1^{\mathrm{r}} = \rmi\left[S_x,S^2\right]
\label{a1r}\end{equation}
for the noncommuting part of $A_1$. After substituting (\ref{a1r}) into (\ref{l1}), we have
\begin{equation}
\rmi \partial_xA_1^{\mathrm{s}} + \left[S^2, S_x\right]_x + \rmi S_{t} = 0.
\label{l1det}\end{equation}
In view of the equalities
\[\pi_{S} \left[S^2, S_x\right]_x  = \left[S^2, S_x\right]_x,\qquad \pi_{S} S_{t} = S_{t}\]
we see that (\ref{l1det}) splits into
\[(\rmid - \pi_{S})\partial_xA_1^{\mathrm{s}} = 0,\qquad
\rmi S_t + \rmi\pi_{S}(\partial_xA_1^{\mathrm{s}}) + \left[S^2, S_x\right]_x = 0, \]
where "$\rmid$" stands for the identity operator. If we set $A_1^{\mathrm{s}} = 0$ (or any other constant matrix
belonging to the centralizer of $S$) we finally derive the matrix NEE 
\begin{equation}
\rmi S_t + \left[S^2, S_x\right]_x = 0
\label{ghf_mat}\end{equation}
written in covariant form. This equation generalizes the usual Heisenberg ferromagnet equation (\ref{hf}) in terms
of Lax pair form and this is why (\ref{ghf_mat}) will further be referred to as generalized Heisenberg equation (GHF).

\begin{rem}
Equation (\ref{ghf_mat}) is a special case of the matrix equation
\begin{equation}
\rmi S_t - a [S,S_{xx}] - \frac{3a}{2} \left(S^2S_{x}S\right)_x + b\left[S^2, S_{x}\right]_x = 0,
\qquad a,b\in\bbc ,
\label{ghf_hom}
\end{equation}
where $S(x,t)\in\mathfrak{sl}(m)$ obeys condition (\ref{s_cond}) but does not obey (\ref{h_red_comp}).
This more general equation is $S$-integrable\footnote{$S$-integrability means integrability in the sense of
inverse scattering transform (S = "scattering") and it has nothing to do with the notation used in the paper
for the matrix-valued function solving the NEEs under consideration.} too and its Lax pair reads 
\begin{eqnarray*}
L(\lambda) &=& \rmi\partial_x - \lambda S,\qquad \lambda\in\bbc,\\
A(\lambda) &=& \rmi\partial_t +  \lambda A_{1} + \lambda^2 A_{2},\qquad
A_{2} = a S + b \left(\frac{2r}{m}I_{m} - S^2\right),\\
A_{1} &=& \rmi a [S,S_{x}] + \frac{3\rmi a}{2}S^2S_{x}S + \rmi b \left[S_{x}, S^2\right],
\qquad r = \frac{1}{2}\tr S^2.
\end{eqnarray*}
After imposing the condition (\ref{h_red}) on the Lax pair above and setting $b=1$, we immediately obtain
(\ref{ghf_mat}). On the other hand, if the matrix $S(x,t)$ can be inverted then the constraint (\ref{s_cond})
implies that $S^2  = I_{m}$ and for $a=2$ one easily derives the usual Heisenberg ferromagnet equation
\[\rmi S_t = \frac{1}{2}[S,S_{xx}].\]
\label{ghfhom_rem}\end{rem}

In view of (\ref{h_red_comp}) it is always possible to represent $S(x,t)$ in the following way
\begin{equation}
S(x,t) = \left(\begin{array}{cc}
0 & \mathbf{u}^T(x,t) \\ \mathbf{v}(x,t) & 0
\end{array}\right),\qquad \mathbf{u}(x,t),\mathbf{v}(x,t)\in M_{n,m}.
\label{s_repr}\end{equation}
Using this representation of $S$, the matrix NEE can be written down "in components" as follows
\begin{equation}
\begin{split}
\rmi\mathbf{u}_t + \left(\mathbf{u}_x\mathbf{v}^T\mathbf{u} - \mathbf{u}\mathbf{v}^T\mathbf{u}_x\right)_x &= 0,\\
\rmi\mathbf{v}_t - \left(\mathbf{v}_x\mathbf{u}^T\mathbf{v} - \mathbf{v}\mathbf{u}^T\mathbf{v}_x\right)_x &= 0 .
\end{split}
\label{ghf_comp}\end{equation}
Important special cases of this system of matrix equations can be obtained when (\ref{s_cond}) is replaced by
some stronger condition. For example, when $m<n$ we can impose the constraint
\begin{equation}
\mathbf{u}^T\mathbf{v} = I_{m}.
\label{uv_con}\end{equation}
Then (\ref{ghf_comp}) is simplified to	
\begin{equation}
\begin{split}
\rmi\mathbf{u}_t + \left(\mathbf{u}_x + \mathbf{u}\mathbf{v}^T_x\mathbf{u}
\right)_x & = 0,\\
\rmi\mathbf{v}_t - \left(\mathbf{v}_x + \mathbf{v}\mathbf{u}^T_x\mathbf{v}\right)_x & = 0.
\end{split}
\label{ghf_rec1}\end{equation}
Similarly, when $m>n$ we can require
\begin{equation}
\mathbf{v}\mathbf{u}^T = I_{n}
\label{uv_con2}\end{equation}
and (\ref{ghf_comp}) turns into	
\begin{equation}
\begin{split}
\rmi\mathbf{u}_t -\left(\mathbf{u}_x + \mathbf{u}\mathbf{v}^T_x\mathbf{u}
\right)_x & = 0,\\
\rmi\mathbf{v}_t + \left(\mathbf{v}_x + \mathbf{v}\mathbf{u}^T_x\mathbf{v}\right)_x & = 0.
\end{split}
\label{ghf_rec2}\end{equation}
It is immediately seen that the latter system can formally be obtained from (\ref{ghf_rec1}) by
inverting the time flow.

\section{Reductions of the GHF}\label{reds}

Here we briefly remind the concept of reductions and reduction group and present a few
simple examples of reductions of the generic matrix system (\ref{ghf_comp}). For those
readers who wish to acquire deeper knowledge on the issues discussed below, we recommend
\cite{mikh1,mikh2,tih}. 

The invariance condition (\ref{h_red}) can be viewed as a result of certain action of $\bbz_2$
onto the set of fundamental solutions of the auxiliary linear (scattering) problem
\begin{equation}
L(\lambda)\Psi(x,t,\lambda) = 0.
\label{specpr}\end{equation}
We shall denote this set of solutions by $\mathcal{D}$. Let us we require that the transformation 
\begin{equation}
\Psi(x,t,\lambda)\to \tilde{\Psi}(x,t,\lambda) = H\Psi(x,t,-\lambda)H, \qquad 
H = \diag(-I_m,I_n) 
\label{red1}
\end{equation}
leaves (\ref{specpr}) intact for any $\Psi(x,t,\lambda)\in\mathcal{D}$. Then it is easily seen 
that the Lax pair (\ref{lax}) must obey (\ref{h_red}). 

Let us now consider the general picture. Assume we have a finite group $\mathrm{G}_{\rm R}$ acting
on $\mathcal{D}$ in the following way
\begin{equation}
\mathcal{K}_g : \Psi(x,t,\lambda)\to\tilde{\Psi}(x,t,\lambda) =
\mathrm{K}_g[\Psi(\kappa_{g}(x,t),k_g(\lambda))],
\quad g\in\mathrm{G}_{\mathrm{R}}.
\label{defred_fs}
\end{equation}
Above $\kappa_g:\bbr^2\to\bbr^2$ is a smooth mapping, $k_g$ is a conformal mapping in the
complex $\lambda$-plane and $\mathrm{K}_g$ is a group automorphism (in our case this is a group
automorphism of $\mathrm{SL}(m+n)$). Then the $G_{\rm R}$-action induced on the Lax operators
is given by
\[L(\lambda)  \to  \tilde{L}(\lambda) = \mathcal{K}_gL(\lambda)\mathcal{K}^{-1}_g,\qquad
A(\lambda)  \to  \tilde{A}(\lambda) = \mathcal{K}_gA(\lambda)\mathcal{K}^{-1}_g.\]
The form of the induced action leads us to the conclusion that the zero curvature condition
remains intact
\[ [\tilde{L}(\lambda),\tilde{A}(\lambda)]=\mathcal{K}_g[L(\lambda), A(\lambda)]\mathcal{K}^{-1}_g = 0.\]
Since the linear problem (\ref{specpr}) is invariant under (\ref{defred_fs}), the scattering operators
$L(\lambda)$ and $\tilde{L}(\lambda)$ differ by a multiplier. As a result, the number of independent entries
in $L(\lambda)$ (dynamical fields) is reduced hence the name of the group $\mathrm{G}_{\mathrm{R}}$ ---
reduction group. In the particular case when $\kappa_g$ is the identity, $\forall g\in\mathrm{G}_{\mathrm{R}}$
we have {\it a local reduction}, otherwise it is {\it nonlocal}. Let us apply these general concepts to our
generic matrix GHF.

\begin{ex} \textbf{Local Reduction}
	
Assume that $\bbz_2\times\bbz_2$ acts onto $\mathcal{D}$ through (\ref{red1}) along with
\begin{equation}
\Psi(x,t,\lambda) \to \tilde{\Psi}(x,t,\lambda) =
\mathcal{E}\left[\Psi^{\dag}(x,t,\lambda^*)\right]^{-1}\mathcal{E},
\label{red2} \end{equation}
where 
\begin{equation}
\mathcal{E} = \diag(\mathcal{E}_m,\mathcal{E}_n), \qquad \mathcal{E}^2 = I_{m+n}.
\label{E_def}\end{equation}
Above $\mathcal{E}_{m}\in M_{m,m}$ and $\mathcal{E}_{n}\in M_{n,n}$ are diagonal matrices.
From the invariance of the scattering problem under (\ref{red1}) and (\ref{red2}), it follows
that (\ref{h_red_comp}) holds as well as
\begin{equation}
\mathcal{E}S^{\dag}(x,t)\mathcal{E} = S(x,t) ,\qquad \mathcal{E}A_{\sigma}^{\dag}(x,t)\mathcal{E}
= (-1)^{\sigma}A_{\sigma}(x,t),\qquad \sigma = 1,2.
\label{lacoef_red2}
\end{equation}
Using the representation (\ref{s_repr}), we see that $\mathbf{u}$ and $\mathbf{v}$ are interrelated through 
\begin{equation}
\mathbf{v}(x,t) = \mathcal{E}_n\mathbf{u}^*(x,t)\mathcal{E}_m.
\end{equation}
As a result, the matrix system (\ref{ghf_comp}) is reduced to the following single NEE
\begin{equation}
\rmi\mathbf{u}_t + \left(\mathbf{u}_x\mathcal{E}_m\mathbf{u}^{\dag}\mathcal{E}_n\mathbf{u}
- \mathbf{u}\mathcal{E}_m\mathbf{u}^{\dag}\mathcal{E}_n\mathbf{u}_x\right)_x = 0.
\label{ghf_comp1}\end{equation}
This reduction was already considered in \cite{ntades17}. In the special case when $\mathbf{u}(x,t)$ and
$\mathbf{v}(x,t)$ are column vectors ($m=r=1$), we can set $\mathcal{E}_m = 1$ without loss of generality.
We shall also require that at least one diagonal element of $\mathcal{E}_n$ is equal to $1$ so that the
condition (\ref{s_cond}) becomes equivalent to 
\[\mathbf{u}^{\dag}\mathcal{E}_n\mathbf{u} = 1.\]
Therefore the matrix equation (\ref{ghf_comp1}) further simplifies to
\begin{equation}
\rmi\mathbf{u}_t + \left[\mathbf{u}_x + \left(\mathbf{u}^{\dag}_x\mathcal{E}_n\mathbf{u}\right)\mathbf{u}
\right]_x  = 0. \quad\Box
\label{ghf_vec1}\end{equation}
\label{locred}\end{ex}

\begin{ex} \textbf{Nonlocal Reduction}

Let us slightly modify the $\bbz_2\times\bbz_2$ reduction from the previous example by
replacing (\ref{red2}) with
\begin{equation}
\Psi(x,t,\lambda) \to \tilde{\Psi}(x,t,\lambda) = 
\mathcal{E}\left[\Psi^{\dag}(-x,t,-\lambda^*)\right]^{-1}\mathcal{E},
\label{red2a}\end{equation}
where $\mathcal{E}$ is the same as in (\ref{E_def}). As a result of the invariance
the scattering problem (\ref{specpr}) under (\ref{red1}) and (\ref{red2a}) we have that the constraints
\[\mathcal{E}S^{\dag}(-x,t)\mathcal{E} = S(x,t) ,\qquad \mathcal{E}A_{\sigma}^{\dag}(-x,t)\mathcal{E}
= (-1)^{\sigma}A_{\sigma}(x,t),\qquad \sigma = 1,2\]
are valid. Therefore we obtain the modified interrelation 
\[\mathbf{v}(x,t) = \mathcal{E}_n\mathbf{u}^*(-x,t)\mathcal{E}_m\]
between $\mathbf{u}$ and $\mathbf{v}$. This time (\ref{ghf_comp}) is reduced to the nonlocal matrix equation
\begin{equation}
\begin{split}
\rmi\mathbf{u}_t(x,t) + \left[\mathbf{u}_x(x,t)\mathcal{E}_m\mathbf{u}^{\dag}(-x,t)\mathcal{E}_n\mathbf{u}(x,t)
- \mathbf{u}(x,t)\mathcal{E}_m\mathbf{u}^{\dag}(-x,t)\mathcal{E}_n\mathbf{u}_x(x,t)\right]_x  = 0.
\end{split}
\label{ghf_comp2}\end{equation}
Restricting ourselves with the vector case ($m=r=1$), we set $\mathcal{E}_m = 1$ again and assume
at least one diagonal element of $\mathcal{E}_n$ is positive like in the previous example. Then
(\ref{ghf_comp2}) is simplified to the vector equation
\begin{equation}
\rmi\mathbf{u}_t(x,t) + \left[\mathbf{u}_x(x,t) + \left(\mathbf{u}^{\dag}_x(-x,t)\mathcal{E}_n\mathbf{u}(x,t)\right)
\mathbf{u}(x,t)\right]_x  = 0
\label{ghf_vec2}\end{equation}
where the column vector $\mathbf{u}$ satisfies the nonlocal constraint 
\[\mathbf{u}^{\dag}(-x,t)\mathcal{E}_n\mathbf{u}(x,t) = 1.\]
For the simplest nontrivial case of a two component vector $\mathbf{u}(x,t)$ this nonlocal reduction was
discussed in \cite{ntades19}. \, $\Box$
\end{ex}

\section{Integrable Deformations}\label{deform}

In this section we shall consider a local integrable deformation of the generic GHF (\ref{ghf_comp}),
i.e. such that contains $x$-derivatives of the potential $S$ only, and present a few reductions. In order
to ensure that the deformation is integrable too, we shall deform the original Lax pair (\ref{lax}) so that the new
pair will constitute a rational bundle.

Let us introduce rational bundle $L(\lambda)$-$A(\lambda)$ pair in generic form:
\begin{equation}
\begin{split}
L(\lambda) &= \rmi\partial_x - \lambda S_{+}(x,t) - \frac{1}{\lambda}S_{-}(x,t),
\quad S_{\pm}(x,t)\in\mathfrak{sl}(m+n),\quad \lambda\in\bbc\backslash\{0\}, \\
A(\lambda) &= \rmi\partial_t + \sum^{2}_{k=-2}\lambda^kA_k(x,t),
\qquad A_k(x,t)\in\mathfrak{sl}(m+n).
\end{split}
\label{laxrat}\end{equation}
We require that the Lax operators above are related to the symmetric space
$\mathrm{SU}(m+n)\slash\mathrm{S}(\mathrm{U}(m)\times\mathrm{U}(n))$, i.e. the reduction (\ref{h_red})
still holds. The symmetry condition (\ref{h_red}) is now equivalent to
\begin{equation}
HS_{\pm}H = - S_{\pm},\qquad HA_{k}H = (-1)^{k}A_{k},\qquad k = -2,\ldots,2	.
\label{lacoef_red1}
\end{equation}  
As discussed in the previous section, (\ref{lacoef_red1}) is a consequence of the $\bbz_2$-action
(\ref{red1}) onto the set of fundamental solutions of (\ref{specpr}).

Similarly to the linear bundle case, we shall require the matrices $S_{+}(x,t)$ and $S_{-}(x,t)$
fulfil
\begin{equation}
S^3_{\pm} = S_{\pm}.
\label{spm_cond}\end{equation}
Using the representation 
\begin{equation}
S_{\pm} = \left(\begin{array}{cc}
0 & \mathbf{u}^T_{\pm} \\ \mathbf{v}_{\pm} & 0
\end{array}\right)
\label{spm_repr}\end{equation}
of the matrices $S_{\pm}(x,t)$ through $\mathbf{u}_{\pm}(x,t),\mathbf{v}_{\pm}(x,t)\in M_{n_{\pm},m_{\pm}}$,
we can write (\ref{spm_cond}) down "in components" as follows:
\begin{equation}
\mathbf{u}_{\pm}\mathbf{v}^T_{\pm}\mathbf{u}_{\pm} = \mathbf{u}_{\pm},\qquad
\mathbf{v}_{\pm}\mathbf{u}^T_{\pm}\mathbf{v}_{\pm} = \mathbf{v}_{\pm}.
\label{uvpm_cond}\end{equation}
The zero curvature condition of the Lax pair (\ref{laxrat}) yields the recurrence relations
\begin{eqnarray}
&&[S_{+},A_{2}] = 0, \label{lp3}\\
&&\rmi \partial_x A_{2}  - [S_{+}, A_{1}] = 0, \label{lp2}\\
&&\rmi \partial_x A_{1} + \rmi \partial_t S_{+} - [S_{+},A_{0}] - [S_{-},A_{2}] = 0, \label{lp1}\\
&&\rmi\partial_x A_{0} - [S_{+},A_{-1}] - [S_{-},A_{1}] = 0, \label{l0a}\\
&&\rmi\partial_x A_{-1} + \rmi\partial_t S_{-} - [S_{-},A_0] - [S_{+},A_{-2}] = 0, \label{lm1}\\
&&\rmi\partial_x A_{-2}  - [S_{-}, A_{-1}] = 0, \label{lm2}\\
&&[S_{-},A_{-2}] = 0 \label{lm3}.
\end{eqnarray}
The analysis of (\ref{lp3})--(\ref{lm3}) resembles much that in the linear bundle case (for the
relations (\ref{l3})--(\ref{l1})). This is why we shall just sketch the procedure below. 

Applying the same argument like for (\ref{lax}), we may pick up $A_{\pm 2}$ to be
\[A_{\pm 2} = \frac{2r_{\pm}I_{m+n}}{m+n} - S_{\pm}^2, \qquad r_{\pm} = \frac{\tr S_{\pm}^2}{2} \,\cdot\]
After splitting the coefficient $A_{1}$  (resp. $A_{-1}$) into $S_{+}$-commuting term (resp. 
$S_{-}$-commuting term) and some remainder $A_{1}^{\mathrm{r}}$ (resp. $A_{-1}^{\mathrm{r}}$) and
inverting the commutator in (\ref{lp2}) (resp. in (\ref{lm2})), we derive
\begin{equation}
A_{\pm 1}^{\mathrm{r}} = \rmi\left[S_{\pm,x},S_{\pm}^2\right]
\label{apm1_res}\end{equation}  
for the non-commuting parts of $A_{1}$ and $A_{- 1}$. Like we did before, we set the $S_{\pm}$-commuting
part of $A_{\pm 1}$ to be 0.

It remains to find the matrix coefficient $A_{0}$. After substituting (\ref{apm1_res}) into (\ref{l0a}),
we obtain the equation
\[\partial_xA_{0} + \left[S_{+}, \left[S_{-}^2,S_{-,x}\right]\right]
+ \left[S_{-}, \left[S_{+}^2,S_{+,x}\right]\right] = 0.\]
It turns out that the commutators above do not always constitute an exact $x$-derivative. Obtaining a local expression
for $A_{0}$ is possible when $\mathbf{u}_{\pm}(x,t)$ and $\mathbf{v}_{\pm}(x,t)$ are all column vectors,
i.e. we have $r_{\pm} = m_{\pm} = 1$. In this case the constraints (\ref{uvpm_cond}) reduce to
\begin{equation}
\mathbf{u}^T_{\pm}\mathbf{v}_{\pm} = 1
\label{uvpm_vecond}\end{equation}
and the result for $A_0$ reads
\begin{equation}
A_{0} = \left(\begin{array}{cc}
\mathbf{u}^T_{+}\mathbf{v}_{-} + \mathbf{u}^T_{-}\mathbf{v}_{+} & 0 \\
0 & - \mathbf{v}_{+}\mathbf{u}^T_{-} - \mathbf{v}_{-}\mathbf{u}^T_{+}
\end{array}\right).
\label{a0res}\end{equation}
This result implies the interrelation
\begin{equation}
\mathbf{u}^T_{+}\mathbf{v}_{+,x} = \mathbf{u}^T_{-}\mathbf{v}_{-,x}
\label{uvxpm}\end{equation}
holds true. Taking into account (\ref{a0res}), we obtain the generic NEEs 
\begin{equation}
\begin{split}
&\rmi\mathbf{u}_{\pm,t} + \left[\mathbf{u}_{\pm,x} 
+ \left(\mathbf{v}^{T}_{\pm,x}\mathbf{u}_{\pm}\right)\mathbf{u}_{\pm}\right]_x
+ 2\left(\mathbf{u}^{T}_{+}\mathbf{v}_{-}
+ \mathbf{u}^{T}_{-}\mathbf{v}_{+}\right)\mathbf{u}_{\pm} = 0,\\
&\rmi\mathbf{v}_{\pm,t} -  \left[\mathbf{v}_{\pm,x}
+ \left(\mathbf{u}^{T}_{\pm,x}\mathbf{v}_{\pm}\right)\mathbf{v}_{\pm}\right]_x
- 2\left(\mathbf{u}^{T}_{+}\mathbf{v}_{-} + \mathbf{u}^{T}_{-}\mathbf{v}_{+}\right)\mathbf{v}_{\pm} = 0
\end{split}
\label{ghf_def1}\end{equation}
from the recurrence relations (\ref{lp1}) and (\ref{lm1}). Let's consider two simple examples (reductions)
of (\ref{ghf_def1}).

\begin{ex} \textbf{Local Reduction}

The pair (\ref{laxrat}) surely fulfils the condition (\ref{uvxpm}) when one introduces the symmetry transformation
\begin{equation}
\Psi(x,t,\lambda) \to  \tilde{\Psi}(x,t,\lambda) = K\Psi(x,t,1/\lambda)K,	
\label{red3}\end{equation}
on the set $\mathcal{D}$. Above the constant diagonal matrix $K=\diag(1,K_n)$, $K^2 = I_{n+1}$ is required
to differ from the unit matrix $I_{n+1}$. Such possibility was already considered by Golubchik and Sokolov \cite{golsok}.

From the invariance of the auxiliary problem (\ref{specpr}) under (\ref{red3}), it follows that the $L(\lambda)$-
$A(\lambda)$ pair (\ref{laxrat}) satisfies
\begin{eqnarray}
&& KL(1/\lambda)K = L(\lambda),\qquad KA(1/\lambda)K = A(\lambda)\quad \Leftrightarrow\nonumber\\
&& S_{-}(x,t) = KS_{+}(x,t)K,\quad A_{-j}(x,t) = KA_{j}(x,t)K,\quad j = 0,1,2.
\label{lacoef_red3}
\end{eqnarray}
Using the representation (\ref{spm_repr}), from the latter equations we see that $\mathbf{u}$ and
$\mathbf{v}$ are interrelated through 
\begin{equation}
\mathbf{u}_{-}(x,t) = K_{n}\mathbf{u}_{+}(x,t),\qquad \mathbf{v}_{-}(x,t) = K_{n}\mathbf{v}_{+}(x,t).
\label{uvpm_red}\end{equation}
Then the system (\ref{ghf_def1}) reduces to
\begin{equation}
\begin{split}
&\rmi\mathbf{u}_{+,t} + \left[\mathbf{u}_{+,x} + \left(\mathbf{v}^{T}_{+,x}\mathbf{u}_{+}\right)\mathbf{u}_{+}\right]_x
+ 4\left(\mathbf{u}^{T}_{+}K_n\mathbf{v}_{+}\right)\mathbf{u}_{+} = 0,\\
&\rmi\mathbf{v}_{+,t} -  \left[\mathbf{v}_{+,x} + \left(\mathbf{u}^{T}_{+,x}\mathbf{v}_{+}\right)\mathbf{v}_{+}\right]_x
- 4\left(\mathbf{u}^{T}_{+}K_n\mathbf{v}_{+}\right)\mathbf{v}_{+} = 0.
\end{split}
\label{ghf_def2}\end{equation}

Now let us impose another $\bbz_2$ reduction in the form given in (\ref{red2}). The matrix coefficients of
(\ref{laxrat}) obey 
\[\mathcal{E}S_{+}^{\dag}(x,t)\mathcal{E} = S_{+}(x,t),
\qquad \mathcal{E}A_{j}^{\dag}(x,t)\mathcal{E} = A_{j}(x,t),
\qquad j= 0,1,2 \]
and the column vectors $\mathbf{u}_{+}$ and $\mathbf{v}_{+}$ are interrelated through\footnote{We restrict
here with interrelations for coefficients and vectors with nonnegative indices since the rest follow
straight from (\ref{lacoef_red3}) and (\ref{uvpm_red}) respectively.}
\[\mathbf{v}_{+}(x,t) = \mathcal{E}_n\mathbf{u}_{+}^*(x,t).\]
We shall also require that at least one diagonal element of $\mathcal{E}_n$ is positive so that 
(\ref{uvpm_vecond}) becomes
\[\mathbf{u}^{\dag}_{+}\mathcal{E}_n\mathbf{u}_{+} = 1.\]
Therefore the vector system (\ref{ghf_def2}) simplifies to a single vector equation of the form
\begin{equation}
\rmi\mathbf{u}_{+,t} + \left[\mathbf{u}_{+,x} + \left(\mathbf{u}^{\dag}_{+,x}\mathcal{E}_n\mathbf{u}_{+}\right)\mathbf{u}_{+}\right]_x
+ 4\left(\mathbf{u}^{\dag}_{+}K_n\mathcal{E}_n\mathbf{u}_{+}\right)\mathbf{u}_{+} = 0. 
\label{ghf_def3}\end{equation}
This vector equation is an integrable deformation of (\ref{ghf_vec1}). $\Box$ \label{locred_def}
\end{ex}

\begin{ex} \textbf{Nonlocal Reduction}
	
Let the reduction group $\bbz_2\times\bbz_2\times\bbz_2$ acts on $\mathcal{D}$ by (\ref{red2a})
and (\ref{red3}). The invariance the spectral problem under this action implies
\[\mathcal{E}S^{\dag}_{+}(-x,t)\mathcal{E} = S_{+}(x,t) ,\qquad \mathcal{E}A_{j}^{\dag}(-x,t)\mathcal{E}
= (-1)^{j}A_{j}(x,t),\qquad j = 0,1,2\]
along with (\ref{lacoef_red3}). As a result, we get the interrelation 
\[\mathbf{v}_{+}(x,t) = \mathcal{E}_n\mathbf{u}^*_{+}(-x,t)\]
between the vectors $\mathbf{u}_{+}(x,t)$ and $\mathbf{v}_{+}(x,t)$. This time (\ref{ghf_def2}) is reduced
to a nonlocal vector equation, namely
\begin{equation}
\begin{split}
\rmi\mathbf{u}_{+,t}(x,t) + \left[\mathbf{u}_{+,x}(x,t) + \left(\mathbf{u}^{\dag}_{+,x}(-x,t)\mathcal{E}_n\mathbf{u}_{+}(x,t)\right)\mathbf{u}_{+}(x,t)\right]_x\\
+ 4\left(\mathbf{u}^{\dag}_{+}(-x,t)K_n\mathcal{E}_n\mathbf{u}_{+}(x,t)\right)\mathbf{u}_{+}(x,t) = 0
\end{split}
\label{ghf_def4}
\end{equation}
where $\mathbf{u}_{+}$ obeys the nonlocal constraint
\[\mathbf{u}_{+}^{\dag}(-x,t)\mathcal{E}_n\mathbf{u}_{+}(x,t) = 1.\]
This equation represents an integrable deformation of (\ref{ghf_vec2}). $\Box$ 
\end{ex}

\section{Conclusion}

We have considered a generic (nonreduced) matrix system of $1+1$-dimensional NEEs associated with the
symmetric space $\mathrm{SU}(m+n)\slash\mathrm{S}(\mathrm{U}(m)\times\mathrm{U}(n))$, see (\ref{ghf_mat}).
This matrix NEE is $S$-integrable with Lax representation in the form of a linear bundle in pole gauge and
represents a mathematical generalization of the classical $1+1$ dimensional Heisenberg ferromagnet equation.
We have shown that this GHF admits a pseudo-Hermitian local $\bbz_2$ reduction, see (\ref{ghf_comp1}),
and another pseudo-Hermitian $\bbz_2$ reduction that is nonlocal, see (\ref{ghf_comp2}). This way we have
extended results already published in \cite{gmv,ntades17,ntades19}. 

As discussed in Remark \ref{ghfhom_rem}, the matrix equation (\ref{ghf_mat}) can be viewed as a
$\bbz_2$-reduction of an even more general matrix equation, see (\ref{ghf_hom}), which seems
to be an unknown one as well. The latter is related to a homogeneous space and combines equations (\ref{ghf_mat})
and the Heisenberg equation (\ref{hf}).

We have constructed a local deformation of the GHF under consideration. This has been done by deforming
the linear bundle Lax pair of the generalized Heisenberg ferromagnet equation to obtain a rational bundle
pair, see (\ref{laxrat}). It turns out this works for the vector case (i.e rank 1 case) otherwise one gets nonlocal
(integral) terms. We have displayed a few reductions (local and nonlocal) of this deformation, see (\ref{ghf_def3})
and (\ref{ghf_def4}), thus extending results published in \cite{gmv,golsok} for the local reduction case.

\section*{Acknowledgements}

The work has been supported by Grant DN 02--5 of Bulgarian Fund "Scientific Research".


\begin{thebibliography}{99}
\bibitem{ablomus}
Ablowitz M., Musslimani Z., Integrable Nonlocal Nonlinear
Schr\"odinger Equation, {\it Phys. Rev. Lett.} {\bf 110} (2013) 064105(5).
\bibitem{BoPo}
Borovik A., Popkov V.,  Completely Integrable Chains of Spin 1,
Preprint N6 of the Institute of Low Temperatures Physics \& Engineering, USSR
Academy of Sciences, Kharkov 1990, 24 pages (in Russian).
\bibitem{forkul}
Fordy A., Kulish P., Nonlinear Schr\"{o}dinger Equations and Simple Lie
Algebras, {\it Commun. Math. Phys.} {\bf 89} (1983) 427--443.
\bibitem{gerdji}
Gerdjikov V., On Nonlocal Models of Kulish-Sklyanin Type and Generalized Fourier Transforms,
In: Advanced Computing in Industrial Mathematics, Studies in Computational Intelligence, 
{\bf 681} Georgiev K., Todorov M., Georgiev I. (Eds.), Springer, Cham., 2017.
\bibitem{ggi}
Gerdjikov V., Grahovski G., Ivanov R., The N-wave Equations with PT Symmetry,
{\it Theor. Math. Phys.} {\bf 188} (2016) 1305--1321.
\bibitem{gmv}
Gerdjikov V., Mikhailov A, Valchev T., Reductions of Integrable Equations
on A.III-type Symmetric Spaces, {\it J. Phys. A: Math. Theor.} {\bf 43} (2010) 434015.
\bibitem{golsok}
Golubchik I., Sokolov V., Multicomponent Generalization of the Hierarchy
of the Landau-Lifshitz Equation, {\it Theor. Math. Phys.} {\bf 124} (2000) 909--917.
\bibitem{gurpek}
G\"urses M., Pekcan A., Integrable Nonlocal Reductions, In:  Symmetries, Differential
Equations and Applications, V. Kac, P. Olver, P. Winternitz and T. \"Ozer (Eds), Springer
Proceedings in Mathematics \& Statistics {\bf 266}, Springer, Cham. 2018, pp~27--52.
\bibitem{ishim}
Ishimori Y., Multi-vortex Solutions of a Two-dimensional Nonlinear Wave Equation,
{\it Prog. Theor. Phys.} {\bf 72} (1984) 33--37.
\bibitem{mikh1}
Mikhailov A., Reduction in the Integrable Systems. Reduction Groups,
{\it Lett. JETF} {\bf 32} (1979) 187--192.
\bibitem{mikh2}
Mikhailov A., The Reduction Problem and Inverse Scattering Method,
{\it Physica D} {\bf 3} (1981) 73--117.
\bibitem{mmnl}
Myrzakulov R., Mamyrbekova G., Nugmanova G., Lakshmanan M., Integrable (2+1) Dimensional 
Spin Model: Geometric and Gauge Equivalent Counterparts, Solitons and Coherent Structures,
{\it Phys. Lett. A} {\bf 233} (1997) 391--396.
\bibitem{ntades19}
Myrzakulov R., Nugmanova G., Valchev T., Yesmakhanova K., Nonlocal Reductions of a Generalized
Heisenberg Ferromagnet Equation, In: Sixth International Conference NTADES 2019, A. Slavova (Ed),
AIP Conference Proceedings {\bf 2159}, AIP Publishing, 2019, 030037-1 -- 030037-13.
\bibitem{blue-bible}
Takhtadjan L., Faddeev L., \emph{The Hamiltonian Approach to Soliton Theory},
Springer, Berlin 1987.	
\bibitem{tih}
Valchev T., On Mikhailov's Reduction Group, {\it Phys. Lett. A} {\bf 379} (2015) 1877--1880.
\bibitem{ntades17}
Valchev T., Yanovski A., New Reductions of a Matrix Generalized Heisenberg Ferromagnet
Equation, {\it Pliska Stud. Math.} {\bf 29} (2018) 179--188.
\bibitem{yantih1}
Yanovski A., Valchev T., Pseudo-Hermitian Reduction of a Generalized Heisenberg Ferromagnet
Equation. I. Auxiliary System and Fundamental Properties, {\it J. Nonl. Math. Phys.} {\bf 25}
(2018) 324--350.

\end{thebibliography}
\end{document}